\documentclass[11pt]{article}
\usepackage{amssymb}
\usepackage{amsmath}
\usepackage{amscd}
\usepackage{latexsym}

\catcode`@=11
\renewcommand{\section}{\@startsection{section}{1}{0pt}{\medskipamount}
{\medskipamount}{\large\bf}}
\numberwithin{equation}{section}
\catcode`@=12
\def\a{\alpha}
\def\b{\beta}
\def\g{\gamma}
\def\de{\delta}
\def\eps{\epsilon}

\def\h{\eta}

\def\la{\lambda}

\def\j{\psi}

\def\Ga{\Gamma}

\def\tG{{\widetilde G}}
\def\tP{{\widetilde P}}
\def\tT{{\widetilde T}}
\def\bT{{T^{\phantom{\dagger}}_\bot}}
\def\bTd{{T^{\dagger}_\bot}}

\newcommand{\gb}{{\bar{g}}}

\newcommand{\zb}{{\bar{z}}}
\newcommand{\bb}{{\bar{\b}}}
\newcommand{\eb}{{\bar{\eta}}}
\newcommand{\jb}{{\bar{\psi}}}
\newcommand{\Db}{{\bar{D}}}

\newcommand{\C}{\mathbb C}
\newcommand{\R}{\mathbb R}
\newcommand{\Z}{\mathbb Z}
\newcommand{\unity}{\boldsymbol{1}}
\newcommand{\Hcal}{{\cal H}}
\newcommand{\Ncal}{{\cal N}}

\def\im{\textrm{i}}
\def\ep{\textrm{e}}
\def\N2{$N{=}2$}
\def\pa{\partial}
\def\pab{\bar\partial}
\def\diff{\textrm{d}}
\def\tr{\textrm{tr}}
\def\str{\textrm{str}}
\def\sfrac#1#2{{\textstyle\frac{#1}{#2}}}
\def\>{\rangle}
\def\<{\langle}
\def\+{\dagger}
\def\={\ =\ }
\def\und{\qquad\textrm{and}\qquad}
\def\tU{\textrm{U}}

\def\be{\begin{equation}}
\def\ee{\end{equation}}
\def\arr{\begin{array}{rll}}
\def\ea{\end{array}}
\def\bea{\begin{eqnarray}}
\def\eea{\end{eqnarray}}

\begin{document}

\begin{titlepage}
\setcounter{page}{0}
\begin{flushright}
CERN-PH-TH/2008-053\\
ITP--UH--05/08\\
\end{flushright}

\vskip 2.0cm

\begin{center}

{\Large\bf  Non-anticommutative Solitons
}

\vspace{15mm}

{\large Sergei V.\ Ketov${}^*$ \ and \ 
Olaf Lechtenfeld${}^{\+\times}$ }
\\[10mm]
\noindent ${}^*${\em Department of Physics,
Tokyo Metropolitan University \\
1-1 Minami-osawa, Hachioji-shi, Tokyo 192-0397, Japan}\\
{Email: ketov@phys.metro-u.ac.jp}
\\[10mm]
\noindent ${}^\+${\em Theory Division, Physics Department, 
CERN, \\
1211 Geneva 23, Switzerland}\\
{Email: olaf.lechtenfeld@cern.ch}
\\[10mm]
\noindent ${}^{\times}${\em Institut f\"ur Theoretische Physik,
Leibniz Universit\"at Hannover \\
Appelstra\ss{}e 2, 30167 Hannover, Germany }\\
{Email: lechtenf@itp.uni-hannover.de}
\vspace{15mm}

\begin{abstract}
\noindent
Certain supersymmetric sigma models in $2{+}1$~dimensions feature
multi-soliton solutions, with and without scattering. We subject 
these systems to a non-anticommutative deformation by replacing the 
Grassmann algebra of the odd superspace coordinates with a Clifford 
algebra. Static $\C P^1$ and scattering U(2) solitons are constructed 
and carry an additional spin-1/2 degree of freedom due to the deformation. 
Abelian BPS solutions exist as well but have infinite action, 
in contrast to the Moyal case.
\end{abstract}

\end{center}
\end{titlepage}

\section{Introduction}
\noindent
Noncommutative solitons occur rather generically in Moyal-deformed
$(2{+}1)$-dimensional field theories 
(see~\cite{Schwarz,DouNe,Szabo} for reviews).
However, they are usually not stable under interaction unless the model
is integrable. In $2{+}1$~dimensions, such is the case for Ward's sigma
model~\cite{Ward88,Ward90}, in which a (Lorentz-breaking) WZW~term insures 
the existence of an underlying linear system. The Ward model appears rather
naturally through a dimensional reduction of the $(2{+}2)$-dimensional
self-dual Yang-Mills system~\cite{MW} to a $(2{+}1)$-dimensional Bogomolny 
system followed by a light-cone gauge fixing, a story which carries over
to the noncommutative realm~\cite{LPS2}. Indeed, Moyal-deformed 
multi-solitons have been constructed and investigated in this setting
\cite{LP3,LP4,Wolf02,ChuLe}.

When supersymmetry is added to the picture, a relation to the twistor
superstring emerges~\cite{Witten03}. Starting from the self-dual 
${\cal N}{\le}4$ super-Yang-Mills equations,\footnote{
For a superspace formulation, see~\cite{KHG}.}
one arrives at a modified U($n$) chiral model with $2{\cal N}{\le}8$
supersymmetries in $(2{+}1)$ dimensions~\cite{Popov07}.
This integrable system may be formulated in a chiral $\R^{2,1|2\Ncal}$ 
superspace and has recently been subjected to a Moyal deformation 
with respect to its bosonic coordinates~\cite{LP8,GIL}.

Yet, noncommutative deformations of superspace naturally involve {\it all\/}
coordinates, i.e.~the fermionic ones as well~\cite{FL1,FL2}.
It is therefore of interest to perform a general Moyal-type deformation
in the aforesaid supersymmetric integrable sigma models and to investigate
their deformed solitons, should they exist. Here, we look at the other extreme,
by replacing the Grassmann algebra of the fermionic chiral-superspace
coordinates by a Clifford algebra. This introduces a non-anticommutative 
supersymmetric sigma model in $(2{+}1)$ dimensions which is still 
integrable.\footnote{
Since we descend from $2{+}2$ dimensions, spinors of opposite chirality
are not related by complex conjugation, and the difficulties encountered
in~\cite{KPT,ChaKu,Cha,InNa,AGVM,JaPu} are avoided.}

The purpose of this paper is the construction and preliminary study of
soliton configurations in such sigma models. After recalling the static
solitons in the conventional ${\cal N}{=}1$ sigma model, we introduce the
non-anticommutative deformation and solve the deformed static BPS equations,  
presenting static classical solutions of nonabelian and abelian type.
Finiteness of the action selects only the former type as a soliton, however.
We then let two such solitons collide by solving the time-dependent
BPS equations for a two-soliton configuration, before closing with comments
and speculations.

\bigskip

\section{Static solitons in the $\Ncal{=}1$ sigma model}
\noindent
We begin with a short review~\cite{per} of soliton configurations in the
$(2{+}0)$-dimensional Grassmannian sigma model with $\Ncal{=}1$~supersymmetry.
The field $\Phi$ lives on the chiral superspace~$\R^{2|2}$ 
coordinatized by~$(z,\zb|\h,\bar\h)$ and takes values in the Grassmannian
Gr$(r,n)=\frac{\tU(n)}{\tU(r){\times}\tU(n{-}r)}$, i.e.
\be
\Phi\in\tU(n) \qquad\textrm{with}\qquad
\Phi^\+\=\Phi \qquad\Leftrightarrow\qquad \Phi^2 \= \unity\ .
\ee
Hence, it may be expressed in terms of a hermitian projector
\be
P \= P^\+ \= P^2\qquad\Leftrightarrow\qquad
P \= T\,(T^\+T)^{-1}T^\+
\ee
based on an $n{\times}r$ matrix function~$T$ via
\be
\Phi \= \unity\ -\ 2P \qquad\Leftrightarrow\qquad
P \= \sfrac12\,(\unity-\Phi) \ .
\ee

The sigma-model equation of motion follows from the action
\be \label{2daction}
S[\Phi] \= \int\!\diff^2z\,\diff^2\h\;\tr\,\Bigl(
(D\Phi)(\Db\Phi) - \sfrac14[\Phi,D\Phi][\Phi,\Db\Phi] \Bigr)\ ,
\ee
where we introduced the superderivatives
\be
 D \=\frac{\pa}{\pa\h} + \im\h\,\frac{\pa}{\pa z} \und
\Db\=-\frac{\pa}{\pa\bar\h} - \im\bar\h\,\frac{\pa}{\pa\zb} \ ,
\ee
which realize the algebra
\be \label{Dalg}
\{D,\Db\} \=0 \qquad\textrm{but}\qquad
D^2 \= \im\pa_z \und \Db^2 \= \im\pa_\zb\ .
\ee
For a soliton, the equation of motion is implied by the BPS condition
\be \label{BPS}
\Db\Phi+\sfrac12[\Phi,\Db\Phi]\=0 \qquad\Leftrightarrow\qquad
(\unity{-}P)\,\Db\,P\=0 \qquad\Leftrightarrow\qquad 
(\unity{-}P)\,\Db\,T\=0\ ,
\ee
which, in view of $PT=T$, is equivalent to
\be
\Db\,T \= T\,\Ga \qquad\quad
\textrm{for some $r{\times}r$ matrix $\Ga(z,\zb|\h,\bar\h)$}\ .
\ee
Since a rescaling $T\to T\Lambda$ does not change $P$, the value of~$\Ga$
is inessential and we may simplify the BPS condition to
\be
\Db\,T \= 0\ ,
\ee
which reveals $T$ as a {\it chiral\/} superfield.
The supersymmetry algebra~(\ref{Dalg}) then implies 
that seperately
\be \label{BPS2}
\frac{\pa}{\pa\bar\h}\,T \=0 \und \frac{\pa}{\pa\zb}\,T \=0
\ee
hold, so that our solutions are given by arbitrary {\it holomorphic\/} 
$n{\times}r$~matrix superfields
\be
T \= R(z|\h) \= R_0(z) + \h\,R_1(z)\ .
\ee

On a BPS configuration, the action evaluates to
\be
S_{\textrm{BPS}}[\Phi] \= 2\int\!\diff^2z\,\diff^2\h\;\tr\,(D\Phi)(\Db\Phi)
\= 8\pi\,|k|
\ee
where $k\in\Z$ is the value of the topological charge
\be
Q \= \sfrac{1}{8\pi} \int\!\diff^2z\,\diff^2\h\;
\tr\,\Bigl( \Phi\,\{ D\Phi,\Db\Phi \} \Bigr)\ .
\ee
Finiteness of the action~(\ref{2daction}) requires $R_0$ and $R_1$ 
to be rational in~$z$, and the largest degree occurring in~$R_0$ 
coincides with~$|k|$. Changing some signs in~(\ref{BPS}) yields an
anti-holomorphic~$T$ and $k<0$, i.e.~anti-solitons.

\bigskip

\section{Non-anticommutative deformation}
\noindent
The above setting has been extended noncommutatively, by conducting 
a Moyal deformation with respect to the bosonic coordinates~\cite{LP8,GIL}.
Replacing the coordinates $(z,\zb)$ by a Heisenberg algebra $[a,a^\+]=1$
not only smoothly deforms the well-known commutative soliton configurations,
but also allows for entirely new, ``abelian'', solitons, which exist even
for $\Phi\in\tU(1)$ and become singular in the commutative limit.
In the noncommutative setting, the BPS solution space is hugely enlarged
because the representation space~$\C^n$ for $\Phi$ and~$P$ gets enhanced
by the Heisenberg-algebra Fock space to~$\C^n{\otimes}\Hcal$. For U(1)~solitons
in particular, the image~$T$ of the projector~$P$ is spanned by coherent 
states~$|\a\>$ of the Heisenberg algebra.

There is no compelling reason in superspace to restrict such a deformation
to the bosonic coordinates. In order to better understand the general case,
we study the other extreme, i.e.~Moyal-deform only the fermionic coordinates.
More concretely, in a general complex superspace $\C^{d|\Ncal}$ we change
the Grassmann algebra of $\{\h_i\}|_{i=1,\dots,\Ncal}$ to the Clifford algebra
\be
\{ \h_i\,,\,\bar\h_j \}\= 2\,C_{ij} \und
\{ \h_i\,,\,\h_j \} \= 0 \= \{ \bar\h_i\,,\,\bar\h_j \} 
\ee
with a hermitian non-anticommutativity matrix~$(C_{ij})$. 
A $\tU(\Ncal)$ rotation diagonalizes~$C$,
and rescalings $(\h_i,\bar{\h}_i)\to(b_i,b^\+_i)$ reduce its eigenvalues
to their signs~$\eps_i\in\{-1,0,+1\}$. Hence, we may write
\be
\{ b_i\,,\,b^\+_j \}\= \eps_i\,\de_{ij} \und
\{ b_i\,,\,b_j \} \= 0 \= \{ b^\+_i\,,\,b^\+_j \} \ ,
\ee
which (when non-degenerate) is $C\ell_{2p,2q}(\R)$ as a real Clifford algebra, 
where $p$ and $q$ are the number of positive and negative eigenvalues of~$C$,
respectively.
In contrast to the bosonic deformation, the representation space of
the Clifford algebra is only finite-dimensional and spanned by the basis
\be
\{\, |s_1,s_2,\ldots,s_\Ncal\> \= 
(b_1^\+)^{s_1} (b_2^\+)^{s_2} \cdots (b_\Ncal^\+)^{s_\Ncal}\,|0\> \,\}
\qquad\textrm{with}\quad s_i\in\{0,1\} \quad\textrm{and}\quad b_i|0\>=0\ ,
\ee
and is endowed with a scalar product of real signature~$(2p,2q)$.
We remark that the $\Ncal{=}2$ ``singlet deformation'' 
$\{b_i,b_j^\+\}=\im\eps_{ij}I$ is equivalent to the case of $C\ell_{2,2}$.

For the rest of the paper we take $\eps_i{=}+1$ and specialize to $\Ncal=1$, 
because only then an action principle is available (see above). 
The Clifford algebra $\{\h,\bar\h\}=C>0$ is rescaled to $\{b,b^\+\}=1$,
which we realize by
\be \label{realize}
1 \ \dot=\ (\begin{smallmatrix} 1 & 0 \\ 0 & 1 \end{smallmatrix}) \quad,\qquad
b \ \dot=\ (\begin{smallmatrix} 0 & 0 \\ 1 & 0 \end{smallmatrix}) \quad,\qquad
b^\+\ \dot=\ (\begin{smallmatrix} 0 & 1\\0 & 0 \end{smallmatrix}) \quad,\qquad
\g := [b,b^\+]
  \ \dot=\ (\begin{smallmatrix}-1 & 0 \\ 0 & 1 \end{smallmatrix}) \ .
\ee
The graded commutation properties with other Grassmann parameters require 
that we adopt an ordering rule for multiplying the above representation 
matrices with (even or odd) scalars~$\la$, namely
\be
b\,\la \ \dot=\ (\begin{smallmatrix} 0 & 0 \\ \la & 0 \end{smallmatrix}) 
\quad,\qquad
\la\,b^\+\ \dot=\ (\begin{smallmatrix} 0 & \la \\ 0 & 0 \end{smallmatrix})
\quad,\qquad
\la\,|0\> \ \dot=\ (\begin{smallmatrix} 0 \\ \la \end{smallmatrix})  
\quad,\qquad 
|1\>\,\la \ \dot=\ (\begin{smallmatrix} \la \\ 0 \end{smallmatrix})
\ee
and the hermitian conjugate relations. Note that $|0\>$ is even while
$|1\>$ is odd.
Multiplication and differentiation acting on bosonic functions~$B$ or
fermionic functions~$F$ are deformed according to the following table
\be \label{map}
\begin{tabular}{|l|cc|cc|cc|}
\hline
Grassmann & 
$\h\,\cdot$ & $\eb\,\cdot$ & $\pa_\h B$ & 
$\pa_\eb B$ & $\pa_\h F$ & $\pa_\eb F$
$\phantom{\Bigm|}$ \\[6pt]
Clifford  & 
$\sqrt{C}\,b\,\cdot$ & $\sqrt{C}\,b^\+\cdot$ &
$\sfrac{1}{\sqrt{C}}[b^\+,B]$ & $\sfrac{1}{\sqrt{C}}[b\,,B]$ & 
$\sfrac{1}{\sqrt{C}}\{b^\+,F\}$ & $\sfrac{1}{\sqrt{C}}\{b\,,F\}$ \\[6pt]
\hline
\end{tabular}
\ee
and the superspace integration turns into the Clifford trace,
\be \label{Clint}
\int\!\diff^2{\h}\;{\cal L} \qquad\longrightarrow\qquad
C^{-1}\str\,{\cal L} \= C^{-1}\tr\,(\g{\cal L}) 
\= C^{-1}\bigl(\<0|{\cal L}|0\>-\<1|{\cal L}|1\>\bigr)\ ,
\ee
which projects onto the $\g$~component of its argument.
As a consequence of the graded cyclic invariance of~$\str$, 
the following rules apply:
\be
\str[B,*] =0= \str\{F,F'\} \quad,\qquad
\str\{B,B'\} = 2\,\str(BB') \quad,\qquad
\str[F,F'] = 2\,\str(FF')\ .
\ee

\bigskip

\section{Deformed solitons}
\noindent
The story of section~2 carries over to the Clifford-deformed realm,
with superfields being replaced by $2{\times}2$ matrix functions with 
diagonal and off-diagonal entries having opposite Grassmann parity.\footnote{
Note that this $2{\times}2$ matrix is tensored with the standard
matrix structure for $\Phi$, $P$ and~$T$.}
In particular, the BPS conditions~(\ref{BPS2}) take the form
\be \label{BPS3}
b\,T \= T\,\Ga \und \pa_\zb\,T \=0\ ,
\ee
where we have reintroduced the freedom of a Grassmann-odd holomorphic 
{\it nilpotent\/} $r{\times}r$~matrix $\Ga(z|b)=\Ga_0(z)+b\,\Ga_1(z)$ 
on the right-hand side. 
Up to rescalings $T\to T\Lambda$, the general solution to~(\ref{BPS3}) reads
\be
T \= R(z|b)\,:\!\ep^{b^\+(\Ga-b)}\!\!: \= 
R(z|b)\,(b\,b^\++b^\+\Ga)\ \ \dot=\ \ R(z|b)\,
\Bigl(\begin{smallmatrix}\Ga_1 & -\Ga_0 \\[4pt] 0 & 1\end{smallmatrix}\Bigr)\ ,
\ee
with a holomorphic polynomial $n{\times}r$ matrix superfield~$R$.
If the factor multiplying~$R$ is invertible it can be dropped,
and we arrive at a smooth Clifford deformation of the well-known
nonabelian soliton configuration of section~2. 

However, when $\Ga_1$ is not invertible, new solutions arise, 
in analogy with the abelian solitons for the bosonic deformation. 
In case $\Ga_1=0$, for example, we get
\be
T \= R(z|b)\;\ep^{b^\+\Ga_0} :\!\ep^{-b^\+b}\!\!: \=
R(z|b)\;\ep^{b^\+\Ga_0}\,|0\>\<0| \ \ \dot=\ \ R(z|b)\,
\Bigl(\begin{smallmatrix} 0 & -\Ga_0 \\[4pt] 0 & 1 \end{smallmatrix}\Bigr)
\qquad\textrm{with}\quad \Ga_0^2=0 \ .
\ee 
In view of the rescaling freedom $T\to T\Lambda$, this is equivalent to taking
$T$ as an $n{\times}r$ matrix of states in the Clifford~module,
\be
|T\> \= R(z|b)\;\ep^{b^\+\Ga_0(z)}\,|0\> \ =:\ R(z|b)\,|\Ga_0\> 
\qquad\textrm{with}\qquad b\,|\Ga_0\> \= |\Ga_0\>\,\Ga_0 \ ,
\ee
where we defined the ($r{\times}r$ matrix of) fermionic coherent 
states~$|\Ga_0\>$. The nilpotency of~$\Ga_0$ is achieved either via its
$r{\times}r$ matrix structure or by the Grassmann-odd nature of its entries.
Generically, the latter is true, and a change of basis in 
$\textrm{image}(P)\simeq\C^r$ diagonalizes the matrix to
\be
\Ga_0(z) \= \textrm{diag}\,\bigl(\b_1(z),\b_2(z),\ldots,\b_r(z)\bigr) 
\qquad\textrm{with}\quad \bigl\{\b_i(z),\b_j(z)\bigr\}=0\ ,
\ee
which yields
\be
|\Ga_0\> \= \textrm{diag}\bigl(\ep^{b^\+\b_i(z)}\bigr)|0\> \=
\textrm{diag}\Bigl( \bigl(1-\b_i(z)b^\+\bigr)\,|0\> \Bigr) \ =:\
\textrm{diag}\bigl( |\b_i\> \bigr)
\qquad\textrm{with}\quad \<\b_i|\b_j\> \= \ep^{\bar\b_i\b_j}
\ee
and simplifies 
\be
|T\>_{ij} \= R_{ij}\bigl(z|\b_j(z)\bigr)\,|\b_j\>
\qquad\textrm{due to}\qquad b|\b_i\> \= \b_i|\b_i\> \ . 
\ee
Note that all states are also holomorphic functions of~$z$.

To be even more concrete, 
let us specialize to $\Phi\in\textrm{Gr}(1,2)=\C P^1$, so that
\be
T_{\textrm{nab}} \= 
\biggl(\begin{matrix} R_1(z|b) \\[4pt] R_2(z|b) \end{matrix}\biggr) \ \simeq\ 
\biggl(\begin{matrix} G(z|b) \\[4pt] 1 \end{matrix}\biggr) \und
|T_{\textrm{ab}}\> \= \biggl(
\begin{matrix} R_1(z|\b) \\[4pt] R_2(z|\b) \end{matrix}\biggr)\,|\b\> \ \simeq\
\biggl(\begin{matrix} G(z|\b) \\[4pt] 1 \end{matrix}\biggr)\,|\b\> \ ,
\ee
where we finally rescaled the lower entry of $T$ to unity.
With these provisions, we can write out the nonabelian projector as
\be \label{Pnab}
P_{\textrm{nab}} \= 
\biggl(\begin{matrix} G\,H\,G^\+ & G\,H \\[4pt] H\,G^\+ & H \end{matrix}\biggr)
\qquad\textrm{with}\qquad H \= (1+G^\+G)^{-1}\ .
\ee
For explicit computations, it is convenient to expand the superfields,
\be
G(z|b) \= g(z) + b\,\j(z) \und
H(z|b) \= h - b\j h^2\gb - gh^2\jb b^\+ - \jb b^\+hb\j + b\j gh^3\gb\jb b^\+\ ,
\ee
where we defined $h=(1+g\gb)^{-1}$. 
For $P_{\textrm{nab}}$ one then obtains the explicit $2{\times}2$ matrix 
\be
\Biggl(\begin{matrix}
gh\gb+b\j h^2\gb+gh^2\jb b^\+-\jb b^\+gh\gb b\j+b\j h^3\jb b^\+ &
gh+b\j h^2-g^2h^2\jb b^\+-\jb b^\+ghb\j-b\j gh^3\jb b^\+ \\[8pt]
h\gb+h^2\jb b^\+-b\j h^2\gb^2-\jb b^\+h\gb b\j-b\j h^3\gb\jb b^\+ &
h-b\j h^2\gb-gh^2\jb b^\+-\jb b^\+hb\j+b\j gh^3\gb\jb b^\+
\end{matrix}\Biggr)
\ee
with operator-valued entries or, via~(\ref{realize}), a $4{\times}4$ matrix
with Grassmann-valued entries, all given in terms of an even and an odd
holomorphic function, namely $g(z)$ and $\j(z)$, respectively.
With this help, one easily computes, for example,
\be
G\,H\,G^\+ \= 1-H + \j h^2\jb \qquad\Rightarrow\qquad 
\str\,\tr\,P_{\textrm{nab}} \= \str\,\bigl(GHG^\++H\bigr) \= 0\ .
\ee

The abelian projector is obtained in a similar fashion,
\be \label{Pab}
P_{\textrm{ab}} \= 
\biggl(\begin{matrix}G\,H\,\bar{G}&G\,H\\[4pt] H\,\bar{G}&H\end{matrix}\biggr)
\ \frac{|\b\>\<\b|}{\<\b|\b\>} \ =:\ P_B \cdot P_F \ ,
\ee
but now with a second odd holomorphic function, $\b(z)$, appearing also in
\be
G(z) \= g(z) + \b(z)\,\j(z) \und
H(z) \= h-\b\,\j h^2\gb-g h^2\jb\,\bb-\b\,\j h^3(1{-}g\gb)\jb\,\bb\ .
\ee
One easily verifies that in this case
\be
G\,H\,\bar{G} \= 1-H \qquad\textrm{so that}\qquad
\str\,\tr\,P_{\textrm{ab}} \= \bigl(\tr\,P_B\bigr)\bigl(\str\,P_F\bigr) \= 1\ .
\ee
Since the moduli $(\b,\bb)$ are Grassmann (and not Clifford), the
matrix~$P_B$ in~(\ref{Pab}) precisely yields the (anti)commutative limit 
of~$P_{\textrm{nab}}$ when replacing $\b\to\h/\sqrt{C}$. Note that this 
differs from the naive replacement $b\to\h/\sqrt{C}$ in~$P_{\textrm{nab}}$ by 
a term $\sim\j\jb\unity$.

The factorized form of~(\ref{Pab}) allows for a U(1)~solution,
\be \label{PU1}
P_{\textrm{U(1)}} \= P_F \= \frac{|\b\>\<\b|}{\<\b|\b\>} \=
b\,(1{+}\b\bb)\,b^\+ + b\,\bb + \b\,b^\+ + \b\,b^\+b\,\bb \ \ \dot=\ \
\biggl(\begin{matrix} \b\bb & \b \\[4pt] \bb & 1+\b\bb \end{matrix}\biggr)\ ,
\ee 
whose bosonic part is necessarily trivial. It depends on a single
odd holomorphic function~$\b(z)$ and is embedded into $\C P^1$
by choosing $G=0$ and $H=1$ in~(\ref{Pab}).

\bigskip

\section{Action and topological charge}
\noindent
We have found plenty of solutions to the Clifford-deformed 
BPS equations~(\ref{BPS3}), and we expect their action to be finite and
equal to the topological charge whenever the coefficient functions
$g(z)$ and $\j(z)$ are polynomial. This claim, however, should be verified
since the action gets deformed to
\be \label{Sdef}
S_{\textrm{BPS}} \= 2\int\!\diff^2z\;\str\;\tr\,\Bigl(
b\,(\pa_z\Phi)\,b^\+(\pa_\zb\Phi) -
\sfrac{\im}{C} \,b\,(\pa_z\Phi)\,[b\,,\Phi] -
\sfrac{\im}{C} [b^\+,\Phi]\,b^\+(\pa_\zb\Phi) -
\sfrac{1}{C^2} [b^\+,\Phi][b\,,\Phi] \Bigr)
\ee
and the topological charge likewise.

Let us begin with the U(1) solution~(\ref{PU1}). Using
\be
[b,P_F] = \b\g-b  \ ,\quad
[b^\+,P_F] = -\bb\g+b^\+ \ ,\quad
b\,\pa_z P_F = -b\,(b^\++\bb)\,\b' \  ,\quad
b^\+\pa_\zb P_F = b^\+(b+\b)\,\bb'
\ee
one quickly finds
\be
S[\Phi_{\textrm{U(1)}}] \= 8\int\!\diff^2z\;
\bigl(\sfrac{1}{C}+\im\bb\b'\bigr)\bigl(\sfrac{1}{C}+\im\b\bb'\bigr)\ .
\ee
The final integral diverges due to the constant term. Therefore,
the U(1) solution cannot be considered as a soliton.\footnote{
Besides, a finite action would depend on the Grassmann-odd moduli~$\b$.}
The same fate hits the abelian solution~(\ref{Pab}): Because
\be
\str\,\Bigl( [b^\+,P_{\textrm{ab}}][b,P_{\textrm{ab}}] \Bigr) \=
\str\,\Bigl( P_B\,[b^\+,P_F]\,P_B\,[b,P_F] \Bigr) \=
P_B^2\;\str\,\Bigl( [b^\+,P_F][b,P_F] \Bigr) \= P_B
\ee
and $\tr\,P_B=1$ we encounter the same infinite constant.

So our hope rests on the nonabelian solution~(\ref{Pnab}).
For the derivatives and commutators in~(\ref{Sdef}) we compute
\bea &&
\pa\biggl(\begin{matrix} GHG^\+ & GH \\[4pt] HG^\+ & H \end{matrix}\biggr) \=
\biggl(\begin{matrix} (1{-}GHG^\+)\pa GHG^\+ & (1{-}GHG^\+)\pa GH \\[4pt]
-HG^\+\pa GHG^\+ & -HG^\+\pa GH \end{matrix}\biggr) \ ,\\[8pt] &&
\pab\biggl(\begin{matrix} GHG^\+ & GH \\[4pt] HG^\+ & H \end{matrix}\biggr) \=
\biggl(\begin{matrix} GH\pab G^\+(1{-}GHG^\+) & -GH\pab G^\+ GH \\[4pt]
H\pab G^\+(1{-}GHG^\+) & -H\pab G^\+ GH \end{matrix}\biggr) \ ,
\eea
where $\pa$ stands for $\pa_z$ or $[b^\+,\cdot]$, and $\pab$ represents
$\pa_\zb$ or $[b,\cdot]$. It is important to recall that
\be
\pa_\zb G \= 0\= [b,G] \und \pa_z G^\+ \= 0 \= [b^\+,G^\+]\ .
\ee
Into the above expressions, 
it is convenient to insert the $2{\times}2$ matrix representations
\be
H \ \dot=\ \biggl(\begin{matrix}
h+\j h^3\jb&-g h^2\jb\\[4pt]-\j h^2\gb&h+\j gh^3 \gb\jb\end{matrix}\biggr) \und
1{-}GHG^\+ \ \dot=\ \biggl(\begin{matrix}
h-\j gh^3\gb\jb&-gh^2\jb\\[4pt]-\j h^2\gb&h-\j h^3\jb\end{matrix}\biggr)
\ee
as well as 
\be
G \ \dot=\ 
\biggl(\begin{matrix} g & 0 \\[4pt] \j & g \end{matrix}\biggr) \quad,\qquad
\pa_z G \ \dot=\ 
\biggl(\begin{matrix} g' & 0 \\[4pt] \j' & g' \end{matrix}\biggr) \quad,\qquad
[b^\+,G] \ \dot=\ 
\biggl(\begin{matrix} \j & 0 \\[4pt] 0 & -\j \end{matrix}\biggr) 
\ee
and their hermitian conjugates. Putting everything together, the
supertrace kills all terms in~(\ref{Sdef}) except for the first one,
which evaluates to
\be \label{Snab}
S[\Phi_{\textrm{nab}}] \= 8\int\!\diff^2z\; \frac{g'\,\gb'}{(1{+}g\gb)^2} 
\ee
independent of~$\j$. For comparison, we have also computed
the deformed topological charge and obtained
\be
Q[\Phi_{\textrm{nab}}] \= 
\sfrac{1}{\pi}\int\!\diff^2z\; \frac{g'\,\gb'}{(1{+}g\gb)^2} \ .
\ee
Thus, for $g(z)$ being polynomial we reproduce the familiar relation
\be
S_{\textrm{BPS}} \= 8\pi\,Q \= 8\pi\,\textrm{deg}(g)\ <\ \infty\ ,
\ee
qualifying $\Phi_{\textrm{nab}}$ as a true soliton.

\bigskip

\section{Adding motion and scattering}
\noindent
To learn more about the properties of the new non-anticommutative solitons,
one would like to study their dynamics. One possibility is to view
the $(2{+}0)$-dimensional Gr$(r,n)$ sigma models with $r=0,1,\ldots,n$
for fixed~$n$ as the static sectors of a $(2{+}1)$-dimensional U($n$)
sigma model, again with $\Ncal{=}1$ supersymmetry. To guarantee the
existence of solitons stable under interaction, we should lift the
integrability to~$\R^{2,1|2}$. This can be achieved by choosing the
(not Lorentz-invariant but integrable) $\Ncal{=}1$ Ward sigma model, 
which descends from the supersymmetric selfdual Yang-Mills system
in $\R^{2,2|2}$~\cite{Popov07,LP8}.

We are not aware of a proper action or energy functional 
for~$\Phi(t,z,\zb|\h,\eb)$ in the supersymmetric Ward model, 
but we can investigate its equations of motion~\cite{LP8,GIL},
\begin{equation} \label{Wardeom}
\begin{aligned}
&(\pa_z{+}\pa_\zb)\bigl(\Phi^\+(\pa_z{+}\pa_\zb)\Phi\bigr) 
-(\pa_z{-}\pa_\zb{+}\im\pa_t)\bigl(\Phi^\+(\pa_z{-}\pa_\zb{-}\im\pa_t)\Phi\bigr)
\= 0 \ ,\\[4pt]
&(\pa_\h{+}\pa_\eb)\bigl(\Phi^\+(\pa_z{+}\pa_\zb)\Phi\bigr)
-(\pa_z{-}\pa_\zb{+}\im\pa_t)\bigl(\Phi^\+(\pa_\h{-}\pa_\eb)\Phi\bigr)
\= 0 \ ,\\[4pt]
&(\pa_\h{+}\pa_\eb)\bigl(\Phi^\+(\pa_z{-}\pa_\zb{-}\im\pa_t)\Phi\bigr)
-(\pa_z{+}\pa_\zb)\bigl(\Phi^\+(\pa_\h{-}\pa_\eb)\Phi\bigr)
\= 0 \ ,\\[4pt]
&(\pa_\h{+}\pa_\eb)\bigl(\Phi^\+(\pa_\h{-}\pa_\eb)\Phi\bigr) \= 0 \ ,
\end{aligned}
\end{equation}
where the unitarity condition reads $\Phi^\+\Phi=\Phi\,\Phi^\+=\unity$.
These equations derive from a linear system, which allows one to employ
the dressing method for constructing time-dependent multi-soliton solutions, 
as was demonstrated in~\cite{LP8}. There it was found that lumps with nonzero 
relative asymptotic velocities keep their speed throughout the evolution
and do not scatter. However, in the limit of coinciding asymptotic velocities
new solutions with typical $\frac{\pi}{2\ell}$ head-on scattering 
emerge~\cite{GIL}.

The above system of equations can be Clifford-deformed by simply applying
the transition~(\ref{map}) to the Grassmann-odd derivatives.
Thereafter, the considerations of~\cite{LP8,GIL} apply to this 
``non-anti\-com\-mu\-ta\-tive Ward model'' almost literally, 
and we may construct a plethora of non-anti\-com\-mu\-ta\-tive 
multi-soliton configurations.
Like in the (anti)commutative setting, a nontrivial scattering behavior
requires taking the coincidence limit for the velocity parameters of the
lumps involved. For the simplest such case, a two-soliton scattering, we may 
choose both lumps to be asymptotically static without loss of generality.
This allows us to utilize our static coordinates $(t,z,\zb|\h,\eb)$ in the
common rest frame and parametrize the time-dependent configuration as
\be
\Phi \= (\unity-2P)\,(\unity-2\tP) \qquad\textrm{with}\qquad
P\=T\,(T^\+T)^{-1}T^\+ \und \tP\=\tT\,(\tT^\+\tT)^{-1}\tT^\+\ .
\ee
This expression fulfils the equations of motion~(\ref{Wardeom}) if
$T$ satisfies the BPS conditions~(\ref{BPS3}) and $\tT$ solves its own
BPS~equations in the background of~$P$~\cite{GIL}, 
\be \label{tBPS}
\pa_\zb\tT + (\pa_\zb P)\,\tT \= 0 \und
\pa_t\tT -2\im(\pa_z P)\,\tT \= 0 \und
[b\,,\tT] + [b{+}b^\+,P]\,\tT \= 0 \ ,
\ee 
which render it non-holomorphic.
Concerning $T$, we take the nonabelian static soliton~$T=R(z|b)$.
For~$\tT$, we copy from~\cite{GIL} the ansatz
\be
\tT \= T\ +\ \bT (\bTd\bT)^{-1}\tG \qquad\textrm{with}\qquad
T^\+\bT \=0 \qquad\Leftrightarrow\qquad \unity-P \= \bT (\bTd\bT)^{-1}\bTd\ ,
\ee
introducing a projector orthogonal to~$P$ 
and a new superfield~$\tG(t,z,\zb|b,b^\+)$. This allows us to simplify the
dressed BPS equations~(\ref{tBPS}) to
\be \label{tBPS2}
\pa_\zb\tG \= 0 \und 
\pa_t\tG -2\im\,\bTd\pa_z T \= 0 \und 
[b,\tG] + \bTd [b^\+,T] \= 0\ .
\ee

Specializing to the U(2)~model, we make use of the scaling freedom to put
\be
T \= \biggl(\begin{matrix} G(z|b) \\[4pt] 1 \end{matrix}\biggr) 
\qquad\Rightarrow\qquad
\bT \= \biggl(\begin{matrix} -1 \\[4pt] G^\+(\zb|b^\+) \end{matrix}\biggr)\ .
\ee
With these inputs, the three equations~(\ref{tBPS2}) become
\be
\pa_\zb\tG \= 0 \und 
\pa_t\tG+2\im\,\pa_z G \= 0 \und 
[b,\tG] - [b^\+,G] \= 0 \ ,
\ee
respectively. Their solution is given by
\be
\tG\= -2\im\bigl(t\,\pa_z G(z|b) + K(z|b)\bigr) - b^\+G(z|b)\,b^\+
\ee
with one more holomorphic superfield~$K$. The last term may be written
as $\ b^\+G\,b^\+=-\sfrac12\bigl[b^\+,[b^\+,G]\bigr]$. Altogether, we have
\be
\tT \= \biggl(\begin{matrix} \ G \ \\[4pt] 1 \end{matrix}\biggr) \ +\
\biggl(\begin{matrix} 1 \\[4pt] \!-G^\+\!\end{matrix}\biggr)\,(1+G\,G^\+)^{-1}
\bigl(2\im\,t\,\pa_z G +2\im\,K + b^\+G\,b^\+\bigr)\ .
\ee

Expanding
\be
G(z|b) \= g(z) + b\,\j(z) \und K(z|b) \= k(z) + b\,\kappa(z)\ ,
\ee
one can work out an explicit matrix expression for~$\tP$ and, 
hence, for~$\Phi$ in terms of the four holomorphic functions above. 
However, this becomes rather lengthy and will not be displayed here. 
Instead, we make some qualitative remarks about the properties of the
constructed two-soliton solution.
Its time dependence, multiplying $\pa_zG$, is linear in~$\tT$ but 
nonpolynomial in~$\Phi$ since $\tT^\+\tT$ must be inverted for computing~$\tP$.
In a $1/t$ expansion for $t\to\pm\infty$, the leading part of~$\tP$ equals
$\unity{-}P$, thus $\Phi$ approaches~$-\unity$ asymptotically.
When we set the nilpotent pieces to zero, the configuration boils down to the
familiar bosonic one~\cite{Ward95,Io,IZ}. Thus, the trajectory of the bosonic 
core is determined by $g$ and~$k$, and for a simple choice like
\be
g(z) \= z \und k(z) \= z^2 
\ee
we already know that $\frac{\pi}{2}$ head-on scattering is produced.

The interesting question is, therefore, in which way the non-anticommuting
degrees of freedom modify the bosonic dynamics. The supersymmetrization gave 
us two additional moduli functions, $\j$ and $\kappa$, and the Clifford 
deformation doubled the representation space by tensoring with~$\C^2$, like 
for a spin-$\frac12$ particle. 
Let us finally take a look at the most extreme case,
\be
G \= b\,\j \und K \= 0\quad,\qquad\textrm{i.e.}\qquad g = k = 0 = \kappa\ .
\ee
It is simple enough to present the soliton configuration explicitly,
\be
\Phi \= \Biggl(\begin{matrix}
1-2\j\jb\,b^\+b+\Sigma_{11}\,b\,b^\+ & -2\j\,b^\++\Sigma_{12}\,b \\[8pt]
-2\jb\,b+\Sigma_{21}\,b^\+ & 1+2\j\jb\,bb^\++\Sigma_{22}\,b^\+b
\end{matrix}\Biggr)
\ee
with
\be 
\begin{aligned}
\Sigma_{11} = -4\im t\j'\jb-4\im t\j\,\jb'-8t^2\j'\jb'(1{+}2\j\jb)\ ,\qquad
\Sigma_{12} = +4\im t\j'(1{+}2\j\jb)-16t^2\j\,\j'\jb'\ ,\\ 
\Sigma_{21} = -4\im t\jb'(1{+}2\j\jb)+16t^2\jb\,\j'\jb'\ ,\qquad
\Sigma_{22} = -4\im t \j'\jb-4\im t\j\,\jb'+8t^2\j'\jb'(1{+}6\j\jb)\ ,
\end{aligned}
\ee
where a prime denotes a derivative with respect to the holomorphic
or antiholomorphic argument. The non-standard large-time behavior
of this configuration arises from
\be
\bigl( 1-\j\,\jb-4t^2\j'\jb'(1{-}\j\jb) \bigr)^{-1} \=
1+\j\,\jb+4t^2\j'\jb'(1{+}\j\jb)
\ee
when inverting $\tT^\+\tT$ and is due to the nilpotent nature 
of the combination $t\j'$.

\bigskip

\section{Conclusions}
\noindent
We have taken a first look at solitons in a non-anticommutative field theory,
namely Ward's integrable U($n$)~sigma model on the deformed superspace
$\R^{2,1|2}_C$ where the index~$C$ indicates a Clifford deformation
$\{\h,\eb\}=2C$ of the fermionic coordinates. We have demonstrated that 
time-dependent multi-soliton configuration with and without scattering can be 
constructed using the familiar dressing method, and displayed a few examples.

In analogy to the bosonic Moyal-deformed situation, new ``abelian'' 
BPS~configurations were found. These are based on fermionic coherent states
and lack a $C{\to}0$~limit. In contrast to their bosonic cousins, however,
their energy turns out to be infinite. Therefore, proper solitons occur
only as non-anticommutative deformations of nonabelian ones. 

For multi-soliton solutions without relative motion of the lumps, 
one is reduced to the static sector, described by a deformed 
$\Ncal{=}1$~Grassmannian sigma model on~$\R^2$. In this situation an
action is known, and its value for the deformed $\C P^1$~solitons was shown 
to agree with their topological charge. The latter remains undeformed and
is determined by the bosonic core alone.

We would like to learn more about the dynamics of these ``non-anticommutative
solitons''. It is interesting to speculate, for instance, that the new
spin-$\sfrac12$ degree of freedom endowed by the Clifford-algebra 
representation space~$\C^2$ might affect their scattering behavior.
However, in the absence of an energy density it is unclear how to extract
physical space-time properties from these field configurations.

\bigskip

\noindent
{\bf Acknowledgements}

\medskip

\noindent
This work was supported jointly by the Deutsche Forschungsgemeinschaft
and the Japan Society for the Promotion of Science.
For hospitality, S.K.~and O.L.~are grateful to the Leibniz University of 
Hannover and to the Tokyo Metropolitan University, respectively.
\bigskip


\begin{thebibliography}{99}
\addtolength{\itemsep}{-6pt}

\bibitem{Schwarz}
  A.~Konechny and A.S.~Schwarz,
  Phys. Rept.  {\bf 360} (2002) 353
  [hep-th/0012145, hep-th/0107251].

\bibitem{DouNe}
  M.R.~Douglas and N.A.~Nekrasov,
  Rev. Mod. Phys.  {\bf 73} (2001) 977
  [hep-th/0106048].

\bibitem{Szabo}
  R.J.~Szabo,
  Phys. Rept.  {\bf 378} (2003) 207
  [hep-th/0109162].

\bibitem{Ward88}
  R.S.~Ward,
  J. Math. Phys. {\bf 29} (1988) 386.

\bibitem{Ward90}
  R.S.~Ward,
  Commun. Math. Phys.  {\bf 128} (1990) 319.

\bibitem{MW}
  L.J.~Mason and N.M.J.~Woodhouse,
  ``Integrability, self-duality, and twistor theory,''\\
  Oxford University Press, Oxford, 1996.

\bibitem{LPS2}
  O.~Lechtenfeld, A.D.~Popov and B.~Spendig,
  JHEP {\bf 06} (2001) 011 
  [hep-th/0103196].

\bibitem{LP3}
  O.~Lechtenfeld and A.D.~Popov,
  JHEP {\bf 11} (2001) 040
  [hep-th/0106213].

\bibitem{LP4}
  O.~Lechtenfeld and A.D.~Popov,
  Phys.\ Lett.\ B {\bf 523} (2001) 178
  [hep-th/0108118].

\bibitem{Wolf02}
  M.~Wolf,
  JHEP {\bf 0206} (2002) 055
  [hep-th/0204185].

\bibitem{ChuLe}
  C.S.~Chu and O.~Lechtenfeld,
  Phys. Lett. B {\bf 625} (2005) 145
  [hep-th/0507062].

\bibitem{Witten03}
  E.~Witten,
  Commun.\ Math.\ Phys.\  {\bf 252} (2004) 189
  [hep-th/0312171].

\bibitem{KHG}
  S.V.~Ketov, H.~Nishino and S.J.J.~Gates,
  Nucl.\ Phys.\  B {\bf 393} (1993) 149
  [hep-th/9207042].

\bibitem{Popov07}
  A.D.~Popov,
  Phys.\ Lett.\  B {\bf 647} (2007) 509
  [hep-th/0702106].

\bibitem{LP8}
  O.~Lechtenfeld and A.D.~Popov,
  JHEP {\bf 0706} (2007) 065
  [arXiv:0704.0530].

\bibitem{GIL}
  C.~Gutschwager, T.A.~Ivanova and O.~Lechtenfeld,
  JHEP {\bf 0711} (2007) 052
  [arXiv:0710.0079].

\bibitem{FL1}
  S.~Ferrara and M.A.~Lled\'o,
  JHEP {\bf 0005} (2000) 008
  [hep-th/0002084].

\bibitem{FL2}
  S.~Ferrara, M.A.~Lled\'o and O.~Maci\'a,
  JHEP {\bf 0309} (2003) 068
  [hep-th/0307039].

\bibitem{KPT}
  D.~Klemm, S.~Penati and L.~Tamassia,
  Class.\ Quant.\ Grav.\  {\bf 20} (2003) 2905
  [hep-th/0104190].

\bibitem{ChaKu}
  B.~Chandrasekhar and A.~Kumar,
  JHEP {\bf 0403} (2004) 013
  [hep-th/0310137].

\bibitem{Cha}
  B.~Chandrasekhar,
  Phys.\ Rev.\  D {\bf 70} (2004) 125003
  [hep-th/0408184].

\bibitem{InNa}
  T.~Inami and H.~Nakajima,
  Prog.\ Theor.\ Phys.\  {\bf 111} (2004) 961
  [hep-th/0402137].

\bibitem{AGVM}
  L.~\'Alvarez-Gaum\'e and M.A.~V\'azquez-Mozo,
  JHEP {\bf 0504} (2005) 007
  [hep-th/0503016].

\bibitem{JaPu}
  I.~Jack and R.~Purdy,
  arXiv:0803.2658.

\bibitem{per}
  A.M.~Perelomov,
  Phys. Rept.  {\bf 174} (1989) 229.

\bibitem{Ward95}
  R.S.~Ward,
  Phys.\ Lett.\  A {\bf 208} (1995) 203.

\bibitem{Io}
  T.A.~Ioannidou,
  J.\ Math.\ Phys.\  {\bf 37} (1996) 3422
  [hep-th/9604126].

\bibitem{IZ}
  T.A.~Ioannidou and W.J.~Zakrzewski,
  J.\ Math.\ Phys.\  {\bf 39}  (1998) 2693
  [hep-th/9802122].

\end{thebibliography}
\end{document}